\documentclass{IEEEtran4PSCC}

\ifCLASSINFOpdf
   \usepackage[pdftex]{graphicx}
\else
   \usepackage[dvips]{graphicx}
\fi

\usepackage[cmex10]{amsmath}

\usepackage[utf8]{inputenc}
\usepackage[margin=1in]{geometry}
\usepackage{booktabs} 

\usepackage{amsmath,amsfonts,amssymb,mathtools}

\usepackage{graphicx}

\usepackage{listings}
\usepackage[ruled,vlined]{algorithm2e}
\usepackage{algorithmic}

\usepackage[colorlinks = true,
            linkcolor = blue,
            urlcolor  = blue,
            citecolor = blue,
            anchorcolor = blue]{hyperref}

\usepackage{tikz}
\usepackage{verbatim}
\usepackage{cite}

\makeatletter
\let\old@ps@headings\ps@headings
\let\old@ps@IEEEtitlepagestyle\ps@IEEEtitlepagestyle
\def\psccfooter#1{%
    \def\ps@headings{%
        \old@ps@headings%
        \def\@oddfoot{\strut\hfill#1\hfill\strut}%
        \def\@evenfoot{\strut\hfill#1\hfill\strut}%
    }%
    \def\ps@IEEEtitlepagestyle{%
        \old@ps@IEEEtitlepagestyle%
        \def\@oddfoot{\strut\hfill#1\hfill\strut}%
        \def\@evenfoot{\strut\hfill#1\hfill\strut}%
    }%
    \ps@headings%
}
\makeatother

\begin{document}

\title{Neuro-physical dynamic load modeling using differentiable parametric optimization}


\author{
\IEEEauthorblockN{Shrirang Abhyankar, J\'an Drgo\v na, Andrew August, Elliott Skomski, Aaron Tuor}
\IEEEauthorblockA{Pacific Northwest National Laboratory (PNNL) \\
Richland, Washington, USA\\
\{shrirang.abhyankar, jan.drgona, andrew.august,  elliot.skomski, aron.tuor\}@pnnl.gov}
}


\maketitle

\begin{abstract}
In this work, we investigate a data-driven approach for obtaining a reduced equivalent load model of distribution systems for electromechanical transient stability analysis. The proposed reduced equivalent is a neuro-physical model comprising of a traditional ZIP load model augmented with a neural network. This neuro-physical model is trained through differentiable programming. We discuss the formulation, modeling details, and training of the proposed model set up as a differential parametric program. The performance and accuracy of this neuro-physical ZIP load model is presented on a medium-scale 350-bus transmission-distribution network.
\end{abstract}

\section{Introduction}

The electrical power grid is going through a transformation of technologies and processes spurred by the deep proliferation of renewable energy sources, smarter control through increasing use of power electronic sources, and availability of high-fidelity measurement sources such as PMUs.  Most of the developments in the electrical grid are emanating from the medium to low-voltage distribution grid. With progressively growing deployment of distributed energy resources (DERs), such as photovoltaics (PV), storage devices, electric vehicles, and microgrids, the characteristics of distribution systems are becoming more complex, both statically and dynamically \cite{heydt2010}. 

A considerable amount of distributed generation is variable and non-dispatchable, primarily comprising of roof-top solar PV panels. The inception of generating sources at the distribution level is not only reducing the net load consumed, but also introducing new patterns of load flow, and voltage abnormalities, including reversed
power flow, which is not well-suited to existing system designs
\cite{kaitraei2011}. Most DERs are interfaced with the grid through power electronic inverters adding complexity of operation due to fast nature of power electronics. The introduction of power electronics is pushing the time-scales of power system transient studies from seconds (rotating generators) to microseconds (power electronic switches). 

Hitherto, the distribution grid has been considered as a ``passive`` network for the transmission network analysis, devoid or having minimal generation resources. With the introduction and increasing penetration of DERs, the distribution grids are becoming much more complex and exhibiting dynamics never seen before. Understanding and faithfully reproducing an accurate response of the distribution system is a big challenge due to the diversity in the distribution system.

\paragraph{Contributions}
We propose a novel neuro-physical reduced equivalent model of distribution systems for electromechanical stability analysis. As opposed to model reduction methods that either lean on physics-based models only or data-driven models, our approach combines these two approaches to realize both their benefits. In this work, the proposed neuro-physical dynamic equivalent comprises of a traditional physics-based ZIP load model augmented with a neural network. This neuro-physical model is set up as a constrained least squares problem and solved through the use of a novel differentiable parametric programming approach that optimizes the physics parameters and the neural network weights together. The neuro-physical model can be used as a surrogate or a reduced-order equivalent of the external area or downstream distribution system to accelerate transient stability simulations of large-scale transmission-distribution grids.

\section{Related Work}

\paragraph{Classical load modeling}
Extensive work on load modeling has been carried out by power system researchers \cite{xiang2016, 43196,son2014improvement,zhang2017dependency} to guide the development of load models.
In the component-based approach~\cite{43196}, an equivalent load model is derived through an aggregation of individual load mix (residential, commercial, and industrial) and load class composition (resistance heating, lighting, room air conditioner, water heating, etc.). 
The measurement-based approach is based on fitting a model on  response characteristics of the distribution system such as the active and reactive power, voltages and three-phase currents. Examples include estimation algorithms based on statistical \cite{5420883,867617} and heuristic techniques \cite{1295013,ma2007measurement,regulski2015estimation}. A typical approach to model loads is through a ZIP load model with its composition (Z, I, P parameters) infered through load composition.
Another line of research work is dynamic equivalencing. This work has been primarily targeted towards finding a reduced equivalent model of external area. There are various techniques used including calculating observable and controllable Grammians, Hankel singular values \cite{zhao2012}, linearized models \cite{Sturk2012,Osipov2018,Zhang2019}, and neural networks \cite{Rehim2012,Ma2012}. 

\paragraph{Differentiable optimization}
Recently, there has been an emergence  on the intersection of constrained optimization and deep learning we could label as \textit{learning to optimize} methods.
Supported problem types range from quadratic programs~\cite{OptNetAmosK17}, stochastic optimization~\cite{DontiAK17}, submodular optimization~\cite{NIPS2017_192fc044}, optimal control problems~\cite{DRGONA202114}, or even combinatorial optimization problems~\cite{Wilder2018}, to name just a few. Reference \cite{DC32021} shows enforcement of hard constraints in learning the solution of constrained nonlinear optimization problems via deep neural networks. More recently, the authors in~\cite{DiffCVxLayers2019} introduced \textit{implicit layers} that are modeled by differentiable constrained optimization solvers and can be used as a plug and play layers in popular deep learning frameworks such as Pytorch or Tensorflow.
The above methods inspired the presented method for load modeling by casting the constrained nonlinear least squares problem as a differentiable parametric program.

\paragraph{Constrained neural architectures}
There are numerous challenges associated with solving constrained  deep learning problems, including guarantees on convergence to stationary points and global minima, smart learning rates, min-max optimization, non-convex regularizers, or constraint satisfaction guarantees~\cite{ConstrainedML2019}. Reference~\cite{ConstrDNN2018} showed how handling of constraints by modifying the Conditional Gradients (CG) algorithm.
Specific neural architectures can be designed to impose  a certain class of hard constraints, such as linear operator constraints~\cite{hendriks2020linearly}.
 Authors in~\cite{logbarrierCNN2019} demonstrated use of a log-barrier method for imposing inequality constraints leading to improved accuracy, constraint satisfaction, and training stability.
Penalty methods based on regularization terms in the loss function have become a popular choice for imposing inequality constraints on the outputs of deep neural network models~\cite{PathakKD15,ConstrCNN7971941}. 
As pointed out by~\cite{MarquezNeilaSF17}, in practice, the existing methods for incorporating hard constraints rarely outperform their soft constraint counterparts despite their weak performance guarantees. 

\section{Transient stability analysis}
Power systems undergo disturbances of various types, including balanced and unbalanced short circuits, outage of generators, transmission lines and other equipment, breaker tripping, etc. \cite{SauerPaiBook}. System planning engineers routinely conduct dynamic security assessment (DSA) studies to analyze the impact of different events, such as a new generator interconnection request or a new equipment installation (e.g. SVCs), or the loss of critical lines or loads. Such studies require the simulation of a lot of what-if scenarios that analyze the system impact of disturbances relative to the base case and different contingency conditions. 

In transient stability analysis, the electrical power system is expressed as a set of nonlinear DAEs, where the differential equations model, $f \in \mathcal{R}^m$,dynamics of the rotating machines (e.g., generators and motors) and the algebraic equations, $g \in \mathcal{R}^n$, represent the nodal balance at the network buses.
\begin{equation}
\begin{aligned}
&\frac{dx}{dt} &=&\:f(x,y),~~~~~p(x^-) \le x \le p(x^+) \\
      &\:0   &=&\:g(x,y)
\end{aligned}
\label{eq:ts1}
\end{equation}

Here, $x$ represents the dynamic or state variable associated with the dynamics of the rotating machines, and $y$ represents the algebraic variables primarily comprising of the network voltages. A typical form of the algebraic equations $g$ is the current-balance form given by (\ref{eq:curr_bal}).
\begin{equation}
    I_{Gen} - I_{Net} - I_{Load} = 0
    \label{eq:curr_bal}
\end{equation}
where, $I_{Gen}$ is the current injection from the generators, $I_{Net}$ is the current transported over the network, $I_{Load}$ is the load consumption. Accurate load modeling plays a very important role in power system transient stability analysis. As transient stability studies are restricted to the high-voltage transmission system (typically 69 kV and up), the aggregated response of the underlying distribution system is modeled through reduced order equivalent load models incident at the distribution substation buses. Faithfully reproducing an accurate response of the distribution system is a big challenge due to the diversity of the loads on the distribution system.

\section{Neuro-physical dynamic load model}
\begin{figure}[h!]
\begin{center}
\includegraphics[scale=0.35]{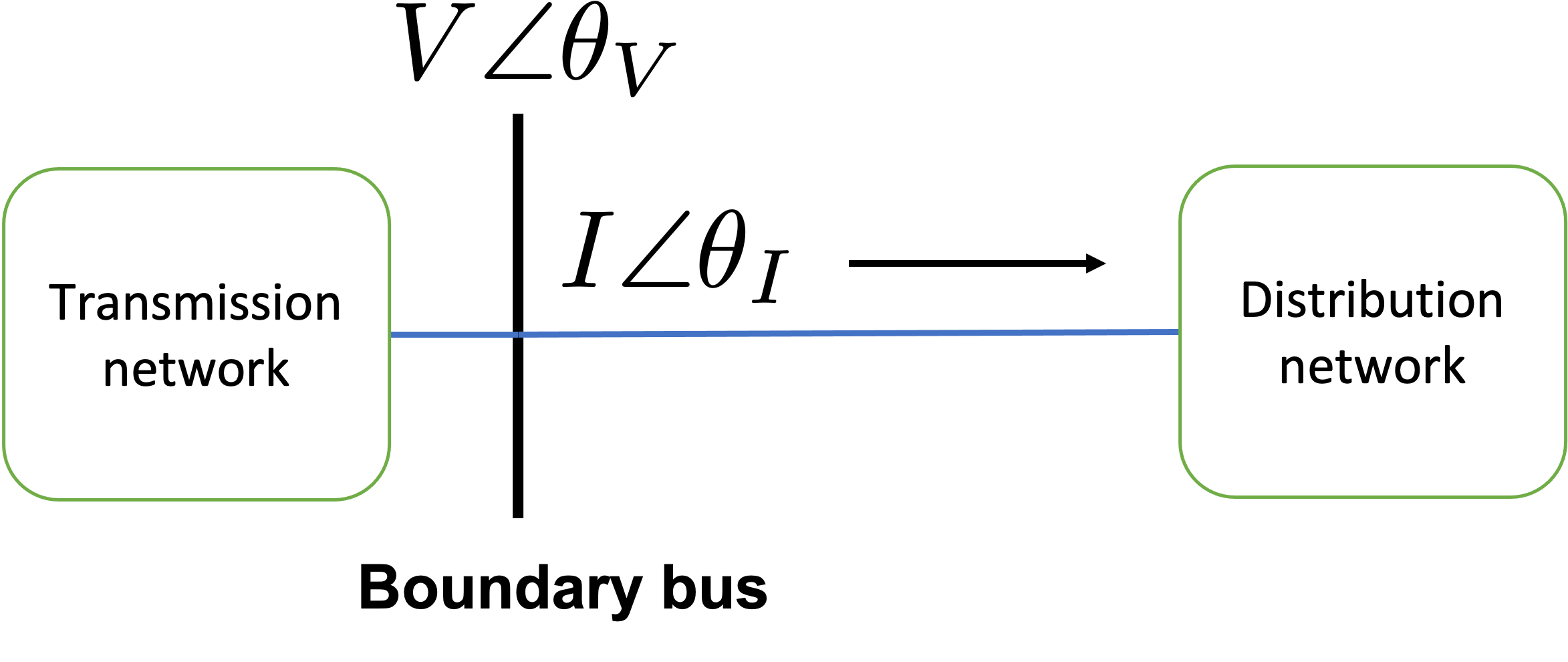}
\caption{Measurements at the transmission-distribution interface}
\label{fig:td_interface}
\end{center}
\end{figure}

The goal of the work is to develop an a reduced equivalent of the distribution network for stability analysis. Figure \ref{fig:td_interface} illustrates the transmission (T) and distribution (D) networks connected at a single boundary bus. We assume presence of measurement devices (such as PMUs) to measure the voltages $\bar{V}(t) = V(t)\angle\theta_V(t)$ and the currents $\bar{I}(t)= I(t)\angle\theta_I(t)$ at the boundary bus.
Using these measurements, an equivalent reduced load model of the external area needs to be discerned given the voltage measurements at the boundary. We assume a following functional form for the power drawn ${S} = f(\bar{V})$, where ${S}$ is the complex power drawn that consists of two components - real power $P$ and reactive power $Q$.

In our work, the proposed neuro-physical load comprises of two components - a physics-based model and a neural network model. We first discuss the ZIP load model which is our choice of physics-based load model used in this paper. Though a ZIP load model is overtly simplistic and may not be an accurate representation of the downstream distribution system, it was solely chosen for implementation ease. Other types of load models, such as motor load models, or composite load models can be also incorporated in the formulation.

\subsection{Physics-based model}
For the physics based model, we will restrict our discussion to the commonly used ZIP load model. This model expresses the real and reactive power drawn $\hat{P}$ and $\hat{Q}$ as a 2\textsuperscript{nd} order polynomial function of the voltages.
 \begin{subequations}
 \label{eq:polynomial}
\begin{align}
\hat{P}(t) &= \alpha_p\hat{P}(t_0) + \alpha_i\hat{P}(t_0)\dfrac{V(t)}{V(t_0)} + \alpha_z\hat{P}(t_0)\dfrac{V(t)}{V(t_0)} ^2 \label{eq:P_physics}\\ 
\hat{Q}(t) &= \beta_p\hat{Q}(t_0) + \beta_i\hat{Q}(t_0)\dfrac{V(t)}{V(t_0)} + \beta_z\hat{Q}(t_0)\dfrac{V(t)}{V(t_0)} ^2 \label{eq:Q_physics}
\end{align}  
 \end{subequations}
The initial load power $\hat{P}(t_0), \hat{Q}(t_0)$ and the initial voltage $V(t_0)$ are known parameters. The coefficients of the load model, $\alpha$ and $\beta$ need to be determined. Typically, this is done through solving the following constrained least-squares problem for $\alpha_p, \alpha_i, \alpha_z$ and $\beta_p,\beta_i,\beta_z$. 
 \begin{subequations}
  \label{eq:least_squares}
\begin{align}
 \min_{\alpha, \beta}  &\sum_{k \in \mathcal{K}}\sum_{t \in \mathcal{T}} (\hat{P}(t,k) - P^{*}(t,k))^2  \\
&+ \sum_{k \in \mathcal{K}}\sum_{t \in \mathcal{T}} (\hat{Q}(t,k) - Q^{*}(t,k))^2 \\
                \text{s.t.} \
                &\alpha_p + \alpha_i + \alpha_z = 1 \\
                &\beta_p + \beta_i + \beta_z = 1 \\
                &\alpha_p, \alpha_i,\alpha_z \ge 0 \\
                &\beta_p, \beta_i,\beta_z \ge 0
\end{align}
 \end{subequations}
Here, $\mathcal{K}$ is the set of time-series trajectories, and $T$ is the number of time steps for each trajectory $k$. The measured power references $P^{*}(t,k)$ and $Q^{*}(t,k)$ at each time-step $t$ is calculated from the voltage and current measurements at the boundary bus.
 \begin{subequations}
\begin{align}
P^{*}(t,k) &= V(t,k)I(t,k)\cos(\theta_V(t,k) - \theta_I(t,k)) \\
Q^{*}(t,k) &= V(t,k)I(t,k)\sin(\theta_V(t,k) - \theta_I(t,k))
\end{align}
 \end{subequations}

\subsection{Neuro-physical ZIP model}
While the physics-based ZIP load model is simplistic, it lacks the fidelity of capturing the complex dynamics primarily because of its restriction to second-order polynomial form. Employing different load models, such as WECC composite load model, is a possibility but it is an extremely complex model and estimating its parameters is a big challenge. In this work, we instead divert to improving the accuracy of a ZIP model using deep neural networks. Thus, the neuro-physical ZIP model combines the benefits of simplicity and some notion of interpretability from the physics-based model along with the expressiveness of the neural network to discover complex dynamics.  

The neuro-physical ZIP model expresses the real $P(t)$ and reactive $Q(t)$ powers as a convex combination of the physics-based  and neural models as follows:
 \begin{subequations}
 \label{eq:convex_comb}
\begin{align}
P_{\text{fit}}(t,k) &= a \hat{P}(t,k) + (1- a) \tilde{P}(t,k) \\
Q_{\text{fit}}(t,k) &= b \hat{Q}(t,k) + (1-b) \tilde{Q}(t,k) 
\end{align}
 \end{subequations}
 Here, $\hat{P}$ and $\hat{Q}$ is the output from the physics-based model (\ref{eq:P_physics}), (\ref{eq:Q_physics}) and $\tilde{P}$ and $\tilde{Q}$ is the output from the neural network model, i.e., $\{ \tilde{P}(t), \tilde{Q}(t) \} = \pi_{\Theta}(V(t), \theta_V(t))$, where $\pi_{\Theta}$ is the neural network model with weights $\Theta$.
 The overall model is given as a convex combination~\eqref{eq:convex_comb} of the polynomial and neural model where $a, b$  are scalars that determine the proportion of the physics-based and neural-network outputs. Depending on their values, the proposed neuro-physical ZIP model can range from a pure physics-based model ($a = 1$, $b = 1$) to an only neural network model ($a = 0$, $b = 0$).

 To train the model~\eqref{eq:convex_comb}  we need to solve the following constrained least squares problem:
 \begin{subequations}
  \label{eq:neuro_aug}
 \begin{align}
 \min_{\Theta, \alpha, \beta}  &\sum_{k \in \mathcal{K}}\sum_{t \in \mathcal{T}} \big( (P_{\text{fit}}(t,k) - P^{*}(t,k))^2 + \\
 & (Q_{\text{fit}}(t,k) - Q^{*}(t,k))^2 \big) \label{eq:ls_loss} \\
                \text{s.t.} \ & \alpha_p + \alpha_i + \alpha_z = 1 \label{eq:con_1} \\
                &\beta_p + \beta_i + \beta_z = 1  \label{eq:con_2} \\
                &\alpha_p, \alpha_i,\alpha_z \ge 0
                \label{eq:con_3} \\
                &\beta_p, \beta_i,\beta_z \ge 0 
                \label{eq:con_4} \\
                & 0 \le a \le 1, \  0 \le  b \le 1
 \label{eq:con_5} 
 \end{align}
 \end{subequations}
 With optimization variables representing the coefficients of the polynomial model $\alpha_p, \alpha_i, \alpha_z$, $\beta_p,\beta_i,\beta_z$, $a$, $b$, and the weights $\Theta$ for the neural network model $\pi_{\Theta}$. 
 
\section{Differentiable Parametric Optimization}
In this work, we leverage  differentiable parametric optimization (DPO) to learn the solution of the original constrained optimization problem.
A generic formulation of the DPO is given as follows:
\begin{subequations}
\label{eq:d3po}
    \begin{align}
 \min_{\Theta} \ & \frac{1}{m} \sum_{i=1}^m { f}({ x^i}, { \xi}^i) & 
 \label{eq:d3po:Q} \\ 
  \text{s.t.} \ &  { g}({ x^i}, { \xi}^i) \leq { 0},   \label{eq:d3po:con1} \\
   \ &  { h}({ x^i}, { \xi}^i) = { 0},   \label{eq:d3po:con2} \\
      \ & { x}^i =   \pi_{\Theta}({ \xi}^i),   \label{eq:d3po:sol} \\
  \ & { \xi}^i \in \Xi, \ \forall i \in \mathbb{N}_1^{m} 
\end{align}
\end{subequations}
Where $\Xi $ represents the sampled dataset, and ${ \xi}^i$ represents $i$-th batch of the sampled problem data.  
The vector $x^i$ represents optimized variables that minimize the loss function while satisfying set of equality~\eqref{eq:d3po:con2} and inequality~\eqref{eq:d3po:con1} constraints.
The mapping~\eqref{eq:d3po:sol} parametrized by $\Theta$ represents the  solution of the underlying constrained optimization problem.
In this work, we use the Neuromancer library~\cite{Neuromancer2021}  for solving the above load modeling given as constrained nonlinear least squares problem~\eqref{eq:neuro_aug}. 
Neuromancer is an open-source Python library built on Pytorch~\cite{paszke2019pytorch} infrastructure that allows 
 to formulate and solve generic DPO problems~\eqref{eq:d3po} by sampling and gradient optimization using AdamW solver~\cite{loshchilov2017decoupled}.

\paragraph{Constraints penalties}
A simple approach to penalize the constraints violations is by augmenting the loss function $\mathcal{L}_{\text{obj}}$~\eqref{eq:d3po:Q} with the penalty functions:
\begin{equation}
\label{eq:ReLU_ineq}
\mathcal{L}_{\text{con}} =  Q_g ||\texttt{ReLU}({ g}({ x}^i, { \xi}^i))||_l + Q_h || { h}({ x}^i, { \xi}^i) ||_l
 \end{equation}
Where $l$ denotes the norm type and
 $Q_g$,  $Q_h$ being the corresponding weight factors. The overall loss then becomes $\mathcal{L} = \mathcal{L}_{\text{obj}} +
\mathcal{L}_{\text{con}}$.

\paragraph{Neuro-physical ZIP model cast as a DPO problem}
To solve the problem~\eqref{eq:neuro_aug}, we cast it in the form~\eqref{eq:d3po}.
Here, the data samples are represented by the power measurements
${ \xi}^k = \{P^{*}(t,k), Q^{*}(t,k) \}$, 
the least square loss~\eqref{eq:ls_loss} is captured by~\eqref{eq:d3po:Q}, the physics-based model given by~\eqref{eq:polynomial},~\eqref{eq:con_1},~\eqref{eq:con_2},~\eqref{eq:con_3},~\eqref{eq:con_4} is compactly represented by a set of equality~\eqref{eq:d3po:con2} and inequality~\eqref{eq:d3po:con1} constraints penalized during training via~\eqref{eq:ReLU_ineq}, While the parametric solution map to be trained~\eqref{eq:d3po:sol} represents the neural model of the problem~\eqref{eq:neuro_aug}.

\section{Case Study}

\subsection{Power system network setup}
\begin{figure}[h!]
\begin{center}
\includegraphics[scale=0.3]{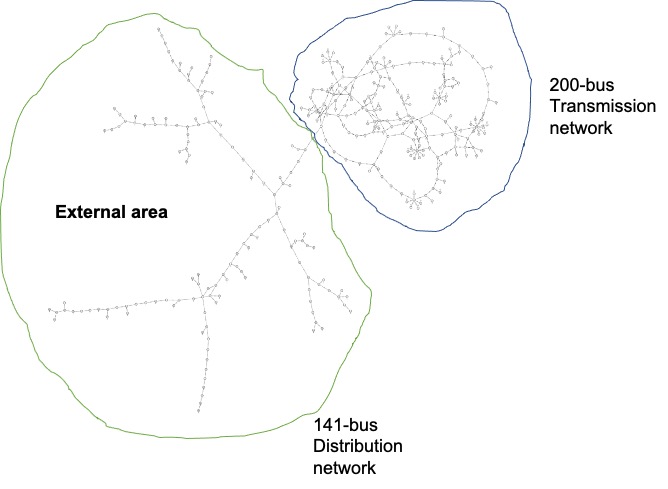}
\caption{Two-area system comprising of a 200-bus transmission network connected to a 141-bus distribution bus at a single boundary node (bus)}
\label{fig:td_interface_2}
\end{center}
\end{figure}

\begin{figure}[h!]
\begin{center}
\includegraphics[scale=0.4]{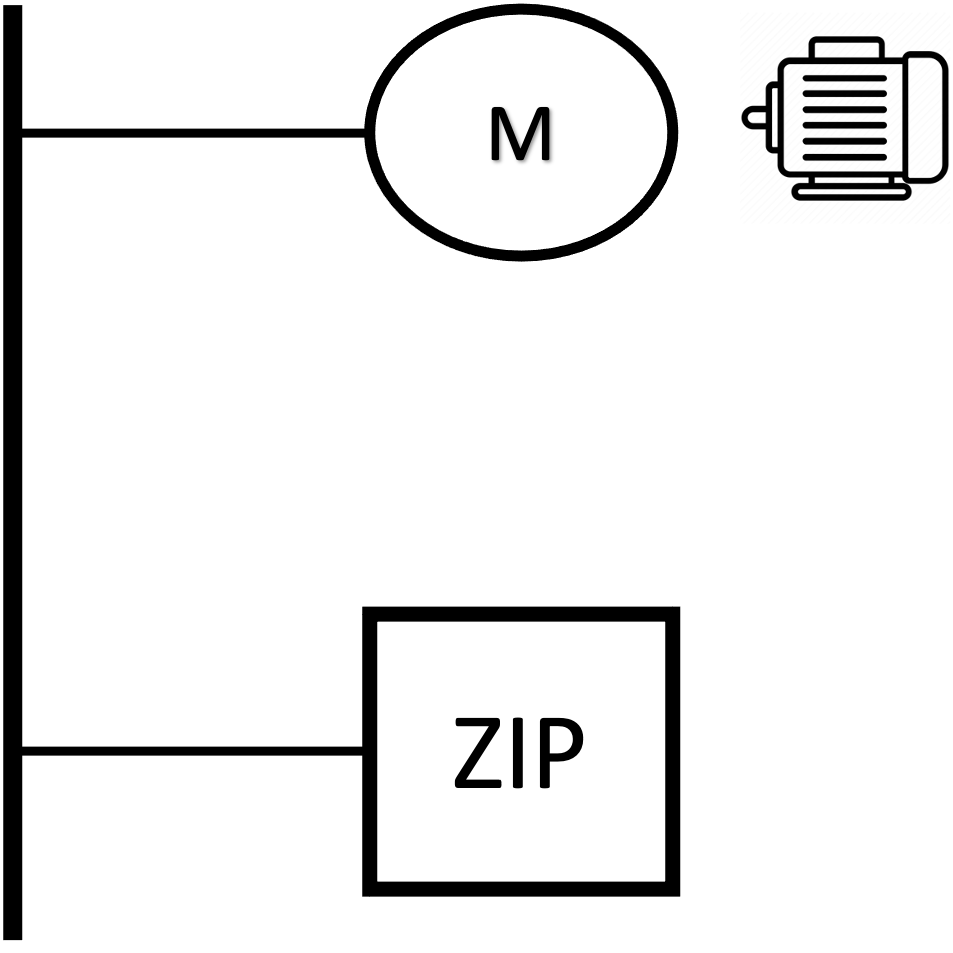}
\caption{Each distribution network node comprises a ZIP load model and three-phase induction motor model}
\label{fig:dist_node}
\end{center}
\end{figure}

The neuro-physical ZIP load model was tested on a synthetic transmission-distribution network made up of 200-bus synthetic network from the ACTIVSg test case repository \cite{activsg1,activsg2,activsg3} connected to a 141-bus distribution network from the MATPOWER \cite{ref-matpower-manual} library. The two networks are connected at a single bus through an added interface line as shown in Fig.~\ref{fig:td_interface_2}.
The 200-bus transmission node has dynamic models of synchronous generators, exciters, and turbine-governors as given in the ACTIVSg test case repository. To incorporate dynamic models in the 141-bus distribution network, each load bus was made up of a combination of a ZIP load model and induction machine motor model as shown in Fig. \ref{fig:dist_node}.
\subsection{Training and test data}
Dynamic simulations of the combined 200-bus transmission+141-bus distribution network were run for three-phase fault on the 141-bus system at different locations. Three scenarios were considered and the following data was collected for training and testing the proposed model.
\begin{itemize}
    \item Three-phase faults at different locations on the distribution grid. Data for 20 trajectories was collected.
    \item Three-phase faults on the transmission side at different locations with the distribution loads modeled as constant impedance. Data for 50 trajectories was collected.
    \item Three-phase faults on the transmission side at different locations with the distribution loads modeled as shown in fig. \ref{fig:dist_node}. Data for 50 trajectories was collected.
\end{itemize}
Each fault simulation was for 10 second total with a fault applied for six cycles (or 0.1 seconds). The voltages at the transmission boundary bus and the current flowing to the distribution network were recorded for each trajectory. 
%

\subsection{Solution setup}
The data for each scenario mentioned above was split into training and test data sets. 60\% of the data, 20\% for validation, and 20\% for testing. We used the open-source library Neuromancer \cite{Neuromancer2021} for solving the problem cast as differential parametric optimization (DPO) problem ~\eqref{eq:d3po}. AdamW~\cite{loshchilov2017decoupled} with learning rate of $0.01$ was used for the optimizer. The relative weights for equality and bounds constraints penalties were set to $Q_g = 1.0$,  $ Q_h = 1.0$. The neural network model we use has four layers with 20 nodes per layer. Overall, models converge to a solution in about 8 mins on CPU.


\subsection{Results and discussion}

\subsubsection{Scenario 1: Three-phase faults at different locations on the distribution grid}
\begin{figure}[h!]
\begin{center}
\includegraphics[scale=0.4, trim = 0cm 0cm 0cm 13cm, clip]{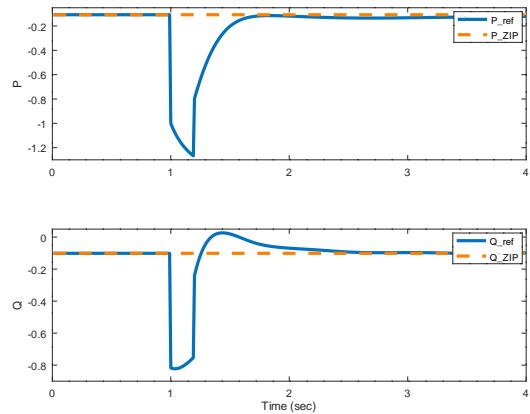}
\caption{Scenario 1: Accuracy of ZIP model for a test trajectory with fault on the distribution side}
\label{fig:solution_200bus_boundary_400zip}
\end{center}
\end{figure}

\begin{figure}[h!]
\begin{center}
\includegraphics[scale=0.5]{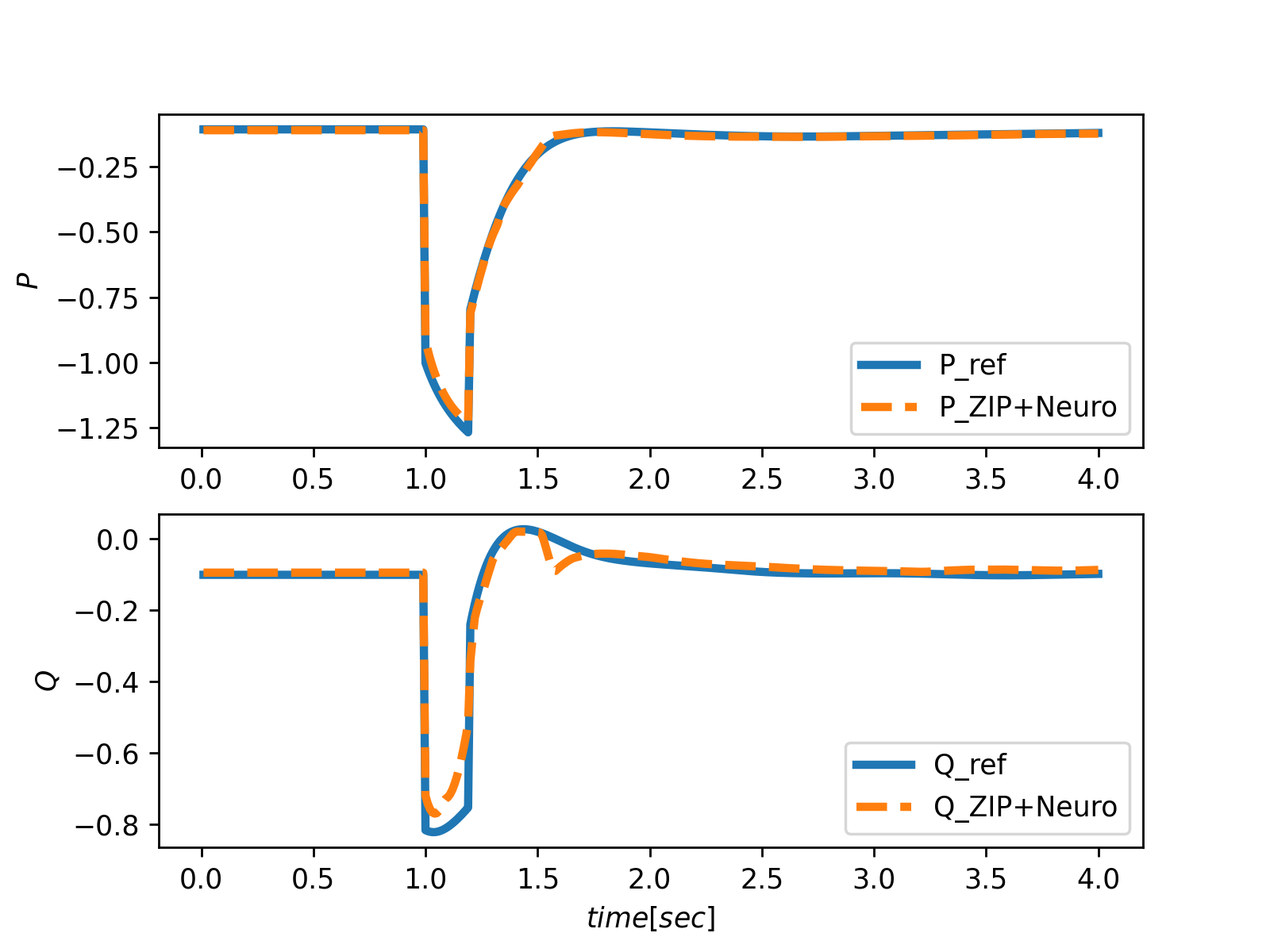}
\caption{Scenario 1: Accuracy of Neuro-augemented ZIP model for a test trajectory with fault on the distribution side}
\label{fig:solution_200bus_boundary_400}
\end{center}
\end{figure}

Figures \ref{fig:solution_200bus_boundary_400zip} and \ref{fig:solution_200bus_boundary_400} shows the comparison of real and reactive power from the ZIP and trained Neuro-ZIP model against the reference trajectories. As seen in the figure, the neuro-ZIP model shows a close match with the reference trajectory, particularly for the real power (P). All the trajectories for scenario 1 resulted in reaching a steady-state after 4 seconds and hence we only used data for the first four seconds of the dynamics simulations.

\subsubsection{Scenario 2: Three-phase faults on the transmission side at different locations with the distribution loads modeled as constant impedance load models} 
\begin{figure}[h!]
\begin{center}
\includegraphics[scale=0.4,trim = 0cm 0cm 0cm 13cm, clip]{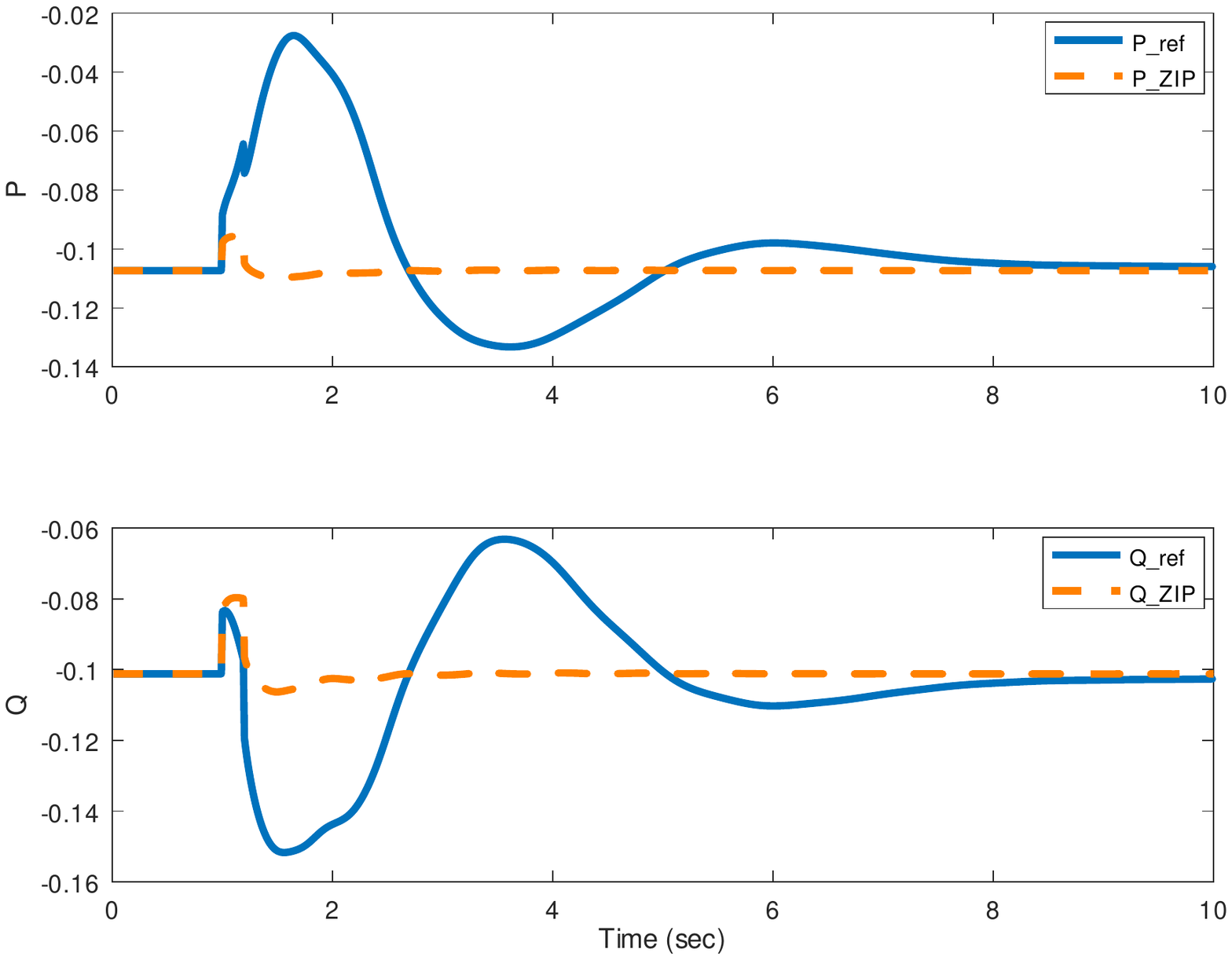}
\caption{Scenario 2: Accuracy of ZIP model for a test trajectory with fault on the transmission side}
\label{fig:solution_200bus_boundary_2zip}
\end{center}
\end{figure}

\begin{figure}[h!]
\begin{center}
\includegraphics[scale=0.5]{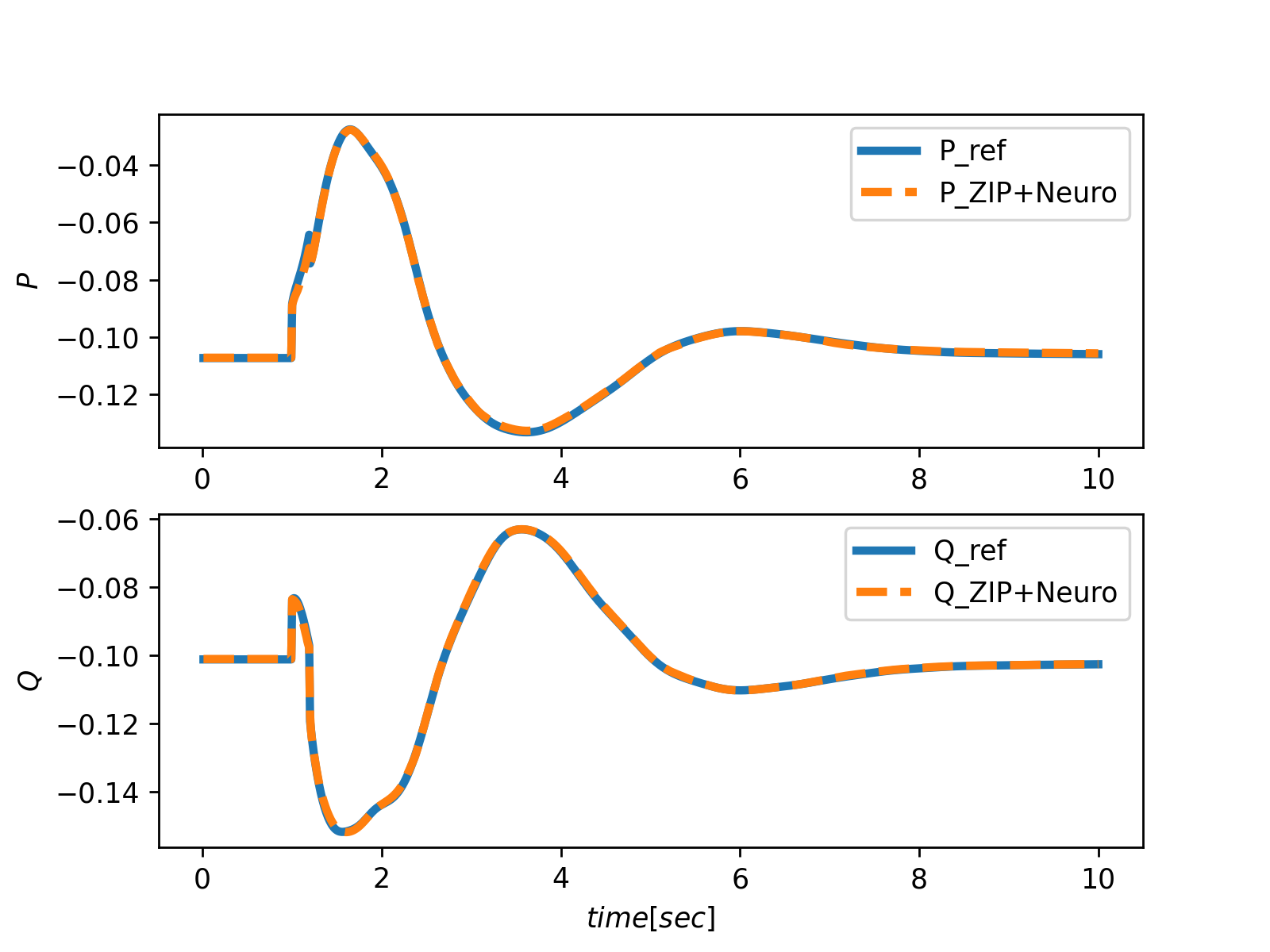}
\caption{Scenario 2: Accuracy of Neuro-augemented ZIP model for a test trajectory with fault on the transmission side}
\label{fig:solution_200bus_boundary_2}
\end{center}
\end{figure}

In this scenario, the faults were applied on the transmission side. The loads at the distribution nodes were modeled as constant impedances. Figures \ref{fig:solution_200bus_boundary_2zip} and  \ref{fig:solution_200bus_boundary_2} compare the performance of the neuro-ZIP model. For this scenario, the proposed model was able to track the reference trajectory very accurately compared to the traditional ZIP model only.

\subsubsection{Scenario 3: Three-phase faults on the transmission side at different locations with the distribution loads modeled as shown in fig. \ref{fig:dist_node}}
\begin{figure}[h!]
\begin{center}
\includegraphics[scale=0.4,trim = 0cm 0cm 0cm 13cm, clip]{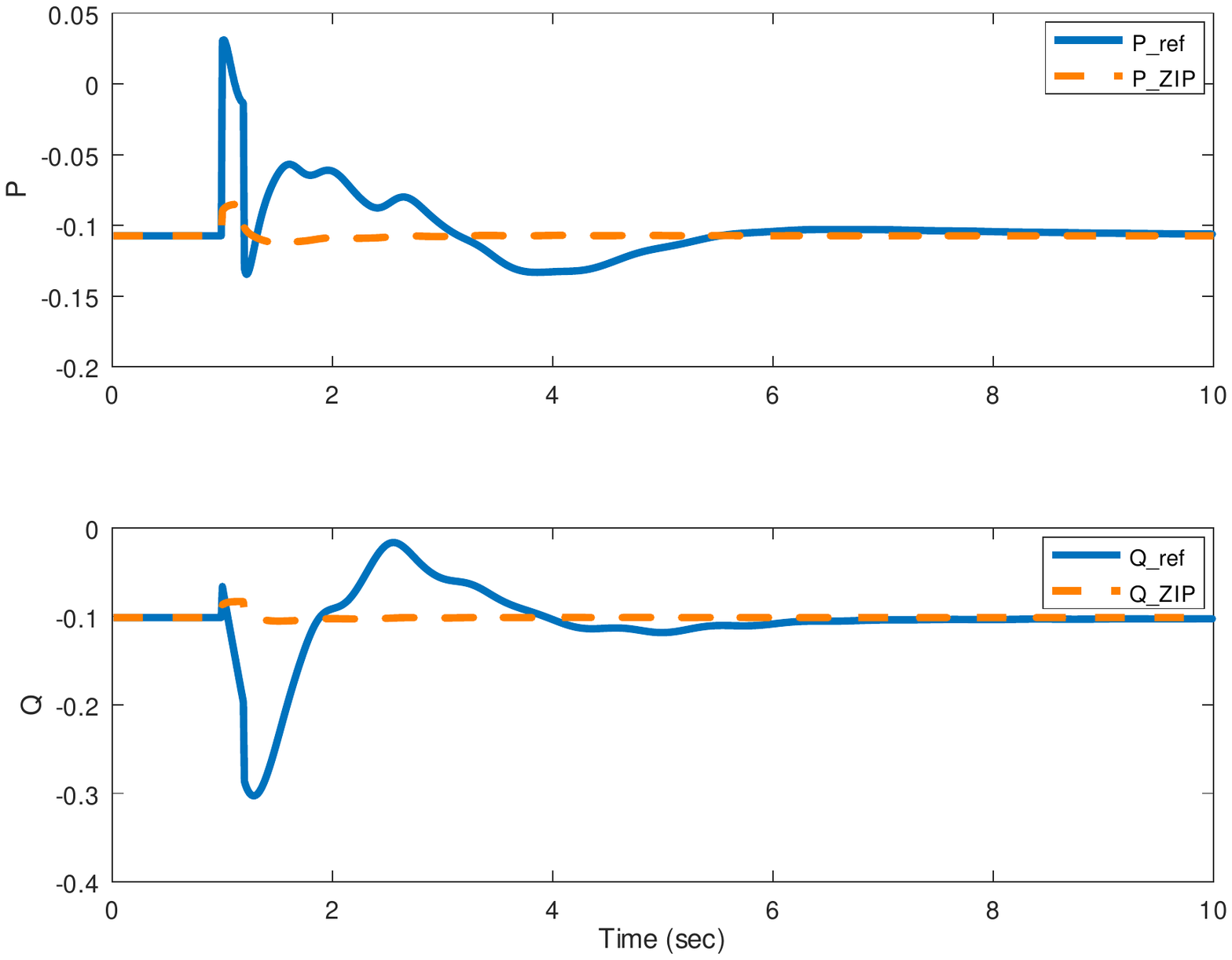}
\caption{Scenario 3: Accuracy of ZIP model for a test trajectory with fault on the transmission side}
\label{fig:solution_200bus_boundary_3zip}
\end{center}
\end{figure}

\begin{figure}[h!]
\begin{center}
\includegraphics[scale=0.5]{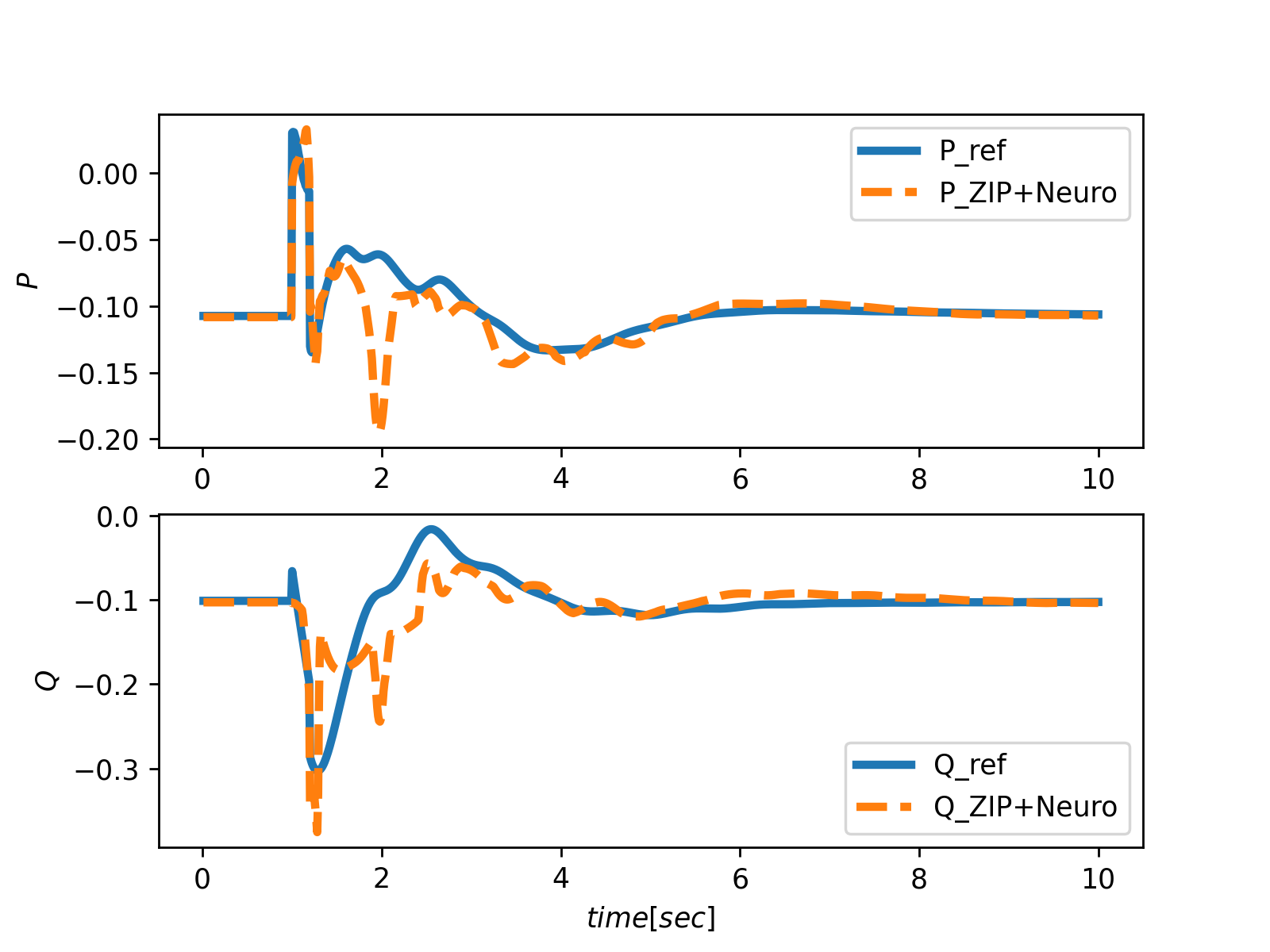}
\caption{Scenario 3: Accuracy of Neuro-augemented ZIP model for a test trajectory with fault on the transmission side}
\label{fig:solution_200bus_boundary_3}
\end{center}
\end{figure}

This scenario is the same as scenario 2 with the exception that the loads on the distribution side are a combination of loads as shown in fig. \ref{fig:dist_node}. Figure \ref{fig:solution_200bus_boundary_3zip} and \ref{fig:solution_200bus_boundary_3} has a comparison of the output from the ZIP and the trained neuro-ZIP model against the reference trajectory. The trained model is able to track the slow moving dynamics to a good degree, but it has some discrepancies in tracking the transient part. 

\begin{table}[h!]
 \centering
    \caption{Prediction accuracy and constraints violations for each scenario}
\begin{tabular}{cccccc}
\toprule
Scenario & MSE $P(t)$   &  MSE $Q(t)$   &  a & b & Con. viol.  \\
\midrule
1 & 0.0004 & 0.0034 & 0.594 & 0.586 & 0.009   \\
2 & 5.05e-06 & 3.32e-07 & 0.591 & 0.594 & 0.009 \\
3 & 0.0006 & 0.0009 & 0.579 & 0.569 & 0.002 \\
\bottomrule
\end{tabular}
    \label{tab:stat_scenario1}
\end{table}

\section{Conclusions and Future Work}
In this paper, we presented a neuro-physical load modeling approach for determining a dynamic load model of distribution grids for transient stability analysis. We formulate this problem as a differentiable constrained nonlinear least-squares problem by learning a convex combination of a  physics-based model and neural network. Our results show the proposed neuro-augmented modeling approach can greatly improve the prediction accuracy of the simplified physics-based model while satisfying the imposed constraints. 

Extensions of the work include finding equivalent of larger networks with multiple boundary buses and use of more complex physics-based load models. The differentiable parametric optimization (DPO) method used in this paper represents a soft-constrained approach for obtaining an approximate parametric solution to the constrained optimization problems. In the future work, employing methods for hard-constraints satisfaction, such as the method presented in~\cite{DC32021}.

\section*{Acknowledgements}
This work was supported through the Data Model Convergence (DMC) initiative at Pacific Northwest National Laboratory (PNNL).




\bibliographystyle{IEEEtran}
\bibliography{bibliography,master}

\vfill
\end{document}